\documentstyle[twoside,fleqn,epsf,espcrc2]{article}
\renewcommand{\floatsep}{9mm plus 2pt minus 0pt}
\textfloatsep=\floatsep
\intextsep=\floatsep

\title{SU(3) glueballs on coarse, anisotropic lattices\thanks{Poster 
session, presented by MP}}

\author{
Colin Morningstar \address{Department of Physics, University of California at 
San Diego, La Jolla, CA 92093-0319} 
and 
Mike Peardon \address{Department of Physics and Astronomy, University of 
Kentucky, Lexington, KY 40506-0055}}

\begin{document}

\begin{abstract}
Results from a calculation of the low-lying glueball spectrum of pure-gauge
SU(3) are used as a test of the effectiveness of improved discretisation
schemes in reducing finite spacing errors. 
Glueball masses are extracted from simulations on anisotropic 
lattices, where the 
temporal lattice spacing is much shorter than the spatial ones. This
allows clearer resolution of the decay of glueball correlators. 
\end{abstract}

\maketitle

\section{INTRODUCTION}
Current interest in the use of Symanzik's improvement scheme, supplemented by
mean-field link renormalisation, to provide a 
better 
discretisation scheme for QCD has stemmed from recent successes of coarse 
lattice simulations \cite{Lepage95}.
Coarse lattices offer the advantage of a
significant reduction in computational overheads, since the approach to the 
continuum limit is accompanied by a rapid increase in CPU time requirements for 
Monte Carlo calculations.
The improvement scheme allowed, for example, calculations of the static
potential at lattice spacings up to $\approx 0.4$ fm with discretisation
errors of only $5\%$. An attempt to use these actions on coarse lattices to 
examine the glueball spectrum of QCD proved difficult, however 
\cite{Morningstar95}. 
The QCD glueballs are massive states and their evaluation on the lattice is
notoriously difficult due to the large vacuum fluctuations of glueball
operators. On a lattice with coarse temporal
discretisation, the glueball correlator can barely be resolved before it is
lost in the noisy vacuum. The use of a large operator basis set to
optimise the ground state overlap was hampered by a technical difficulty caused
by the $2 \times 1$ rectangle in the improved action. This
term couples next-to-nearest neighbouring time-slices and thus modifies the
transfer matrix, making it non-hermitian (physically, this term leads to the
presence of unwanted extra modes in the improved gluon dispersion relation). A 
hermitian, positive transfer matrix is required for the validity of the 
variational
calculation employed and to ensure that the effective mass approaches its
asymptotic value from above. These problems
naturally suggest breaking the
explicit Euclidean symmetry of the action and using a different
lattice spacing in the spatial and temporal directions. With a shorter temporal
lattice spacing, $a_t$, clear evidence for 
plateaux in the effective masses can be seen before the signal is lost.
Once discretisations of this form are used, a solution to the transfer matrix 
problem arises too. 
 
In this article, a study of three of the lighter glueball states
at six different lattice spacings is discussed. For all but the scalar
glueball, the results show the 
expected scaling behaviour.

\section{ANISOTROPIC LATTICE QCD}
The improvement scheme can be extended readily to anisotropic lattices
\cite{Morningstar96}. In this study, the following action was used:
\begin{eqnarray}
{\cal S}_{I\!I}\!\!\!\!&\!=\!&\!\!\!-\beta\!\sum_{x,s>s^\prime}
\frac{a_t}{a_s} \left\{
\frac{5}{3}  \frac{P_{ss^\prime}}{u_s^4}
- \frac{1}{12} \frac{R_{ss^\prime}}{u_s^6}
- \frac{1}{12} \frac{R_{s^\prime s}}{u_s^6}
\right\} \nonumber \\ \label{eq:Action} 
&&\!\!\!-\beta \; \sum_{x,\,s} \;\,
\frac{a_s}{a_t} \left\{
\frac{4}{3}  \frac{P_{st}}{u_s^2 u_t^2}
- \frac{1}{12} \frac{R_{st}}{u_s^4 u_t^2}
\right\},
\end{eqnarray}
with $P_{\mu\nu}$ the plaquette in the $\mu\!-\!\nu$ plane, and $R_{\mu\nu}$
the $2\times1$ $\mu\!-\!\nu$ plane rectangle (with longest side parallel to the
 $\mu$ axis).
This action, intended for use with $a_t \ll a_s$, has $O(a^4_s,a^2_t)$ 
discretisation 
errors. 
The coefficients were determined using tree-level perturbation theory
and tadpole improvement
\cite{Lepage93}. 
The spatial mean link, $u_s$,
was tuned such that the input value in the action matched its measured value as
defined by $u_s=\langle P_{ss'} \rangle^{1/4}$. 
This tuning is rapidly convergent and requires a minimal amount of work.
The temporal mean link, $u_t$, was fixed to $u_t=1$, since its value in
Landau gauge differs from unity by $O(a_t^2/a_s^2)$.

\section{SIMULATION DETAILS}
A set of six simulations were performed at fixed anisotropy, $a_t/a_s = 1/3$, 
with the intention of examining the scaling behaviour of the glueball masses
using this action. The run
parameters are given in Table \ref{tab:RunParams}. Configurations were
generated using the Cabibbo-Marinari (CM) pseudo-heatbath and Creutz 
over-relaxation (OR) methods. 
In our previous coarse lattice calculation
\cite{Morningstar95}, large 
statistical samples were gathered in an attempt to resolve the glueball
correlators. For this work, much smaller statistical samples were required and
the largest data set collected (for the coarsest lattice) consisted of 100
bins, with each bin containing an average of 100 measurements, each measurement
separated by 3 sweeps of a 1-3 hybrid CM/OR update. This is to be contrasted
with the 334 bins of 1000 measurements from the old calculation, which yielded
far less information. 

\subsection{Setting the scale}
The lattice spacing was set by fitting the on-axis static potential to a
standard ansatz, 
\begin{equation}
V(r) = V_0 - \frac{e}{r} + \sigma r,
\end{equation}
then using the fit to determine the hadronic scale, $r_0$ \cite{Sommer94}, 
defined as $r^2 dV/dr = 1.65$ at $r=r_0$,
which corresponds roughly to $r_0 \approx 0.5$fm.
This definition gives a lattice scale that is less dependent on the choice of
potential fit ansatz than the string tension scale. The results of this
determination are given in Table
\ref{tab:RunParams}. In all these calculations, the input ratio $a_t/a_s$ has 
been used since its renormalisation is known to be small \cite{Morningstar96}. 
In all runs, the lattice volumes were chosen 
to be large enough such that the masses of the scalar and tensor states should 
lie within $\frac{1}{2}\%$ of their infinite volume limits
as determined by L\"uscher's finite volume analysis \cite{Luescher86}.

\begin{table}[t] 
\caption{Run parameters for our simulations, $a_t/a_s = 1/3$. The lattice
spacing is set using $r_0=0.53$ fm. Errors in $a_s$ are purely statistical.
\label{tab:RunParams} }
\begin{center}
\begin{tabular}{ccccc}
\hline
$\beta$ & Lattice & $u_{s}$ & $a_s$ (fm) & L(fm)\\
\hline
1.7 & $6^3\times18$ & 0.745 & 0.457(4) & 2.7 \\
1.9 & $6^3\times18$ & 0.764 & 0.414(3) & 2.5 \\ 
2.0 & $8^3\times24$ & 0.772 & 0.387(3) & 3.1 \\
2.2 & $8^3\times24$ & 0.789 & 0.329(2) & 2.6 \\ 
2.4 & $8^3\times24$ & 0.806 & 0.272(4) & 2.2 \\
2.6 & $10^3\times30$ & 0.819 & 0.217(3) & 2.2 \\
\hline
\end{tabular}
\end{center}
\end{table}

\subsection{Optimising the ground state overlap}
The low signal-to-noise ratio of glueball calculations makes it crucial to find
lattice operators with good ground state overlaps. 
Large bases of at least 24 operators were used and optimal correlators were
determined using the variational method. The glueball operators were built 
from spatial Wilson loops
constructed from either APE smeared or Teper fuzzed links \cite{FuzzSmear}. As 
in previous investigations, the smearing and fuzzing
techniques were found to be far more efficient when they included a 
gauge-invariant projection 
of the new links back onto $SU(3)$. In most cases, the APE
smeared operators gave larger contributions to the ground state; 
the glueballs extend over a small number of sites of the
lattice, and thus operators
built from fuzzed ``superlinks'' are too large. At most, one iteration of
Teper fuzzing proved useful. In all cases, these procedures
gave excellent ground state overlaps and in many cases, the overlaps were 
statistically consistent with unity. 

\section{RESULTS}
\begin{figure}[t]
\setlength\epsfxsize{78mm}
\epsfbox{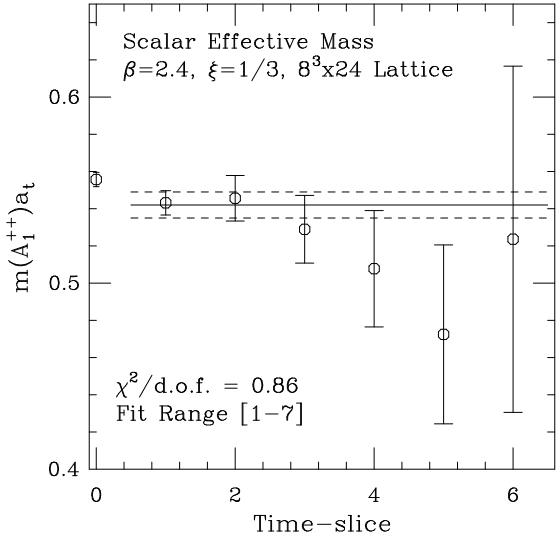}
\vspace{-8ex}
\caption{The effective mass for the scalar glueball at $\beta = 2.4$ 
 ($a_s\approx 0.27$ fm). 
The lines show results from fitting the correlator to
a single exponential.
\label{a1pp_eff_mass}}
\end{figure}
\subsection{The scalar glueball}
The effective mass for the scalar glueball for one of the simulations performed
is shown in Figure \ref{a1pp_eff_mass}. For this example, a correlator 
signal can be
seen out to time-slice 7, and a good single exponential fit 
in the range $1-7$ can be performed. A signal for the first
excited state can be obtained from the first excited variational eigenvector. 
The mass of this state is slightly less than twice
that of the ground state, consistent with an excited state observed in Wilson 
action calculations \cite{Glueballs}. 
The dependence on the lattice spacing is shown in 
Figure \ref{glueball_scale}. The scalar mass shows a significant 
``dip'' as the lattice spacing is increased,
at the bottom of which the glueball is about $70\%$ of its continuum
value (from Wilson data). This dip is caused possibly by the presence of a
critical point in the fundamental/adjoint plane, similar to that which affects
the scalar glueball mass calculated using the Wilson action \cite{Heller96}.
No continuum extrapolations to the scalar data have been included.

\subsection{Other glueball states}
\begin{figure}[t]
\setlength\epsfxsize{78mm}
\epsfbox{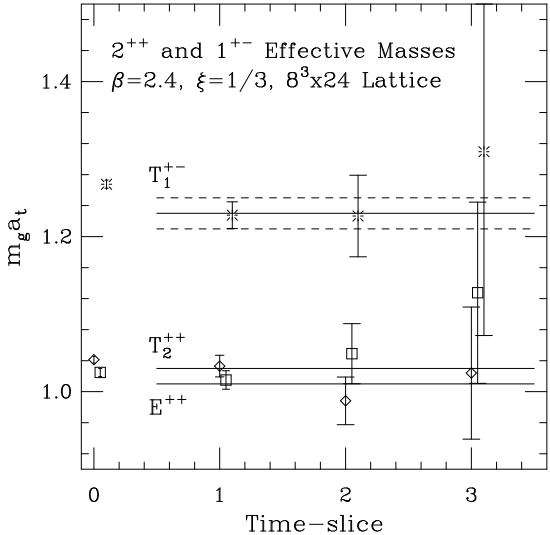}
\vspace{-8ex}
\caption{The effective masses for the tensor and pseudovector states at
$\beta=2.4$. The three irreps $E^{++},T_2^{++} \mbox{ and } T_1^{+-}$ are
plotted as $\Box, \Diamond \mbox{ and } *$, respectively. 
Solid lines show results from fitting correlators to single exponentials.
\label{tens_vec_eff_mass}} 
\end{figure}
\begin{figure}[t]
\setlength\epsfxsize{80mm}
\epsfbox{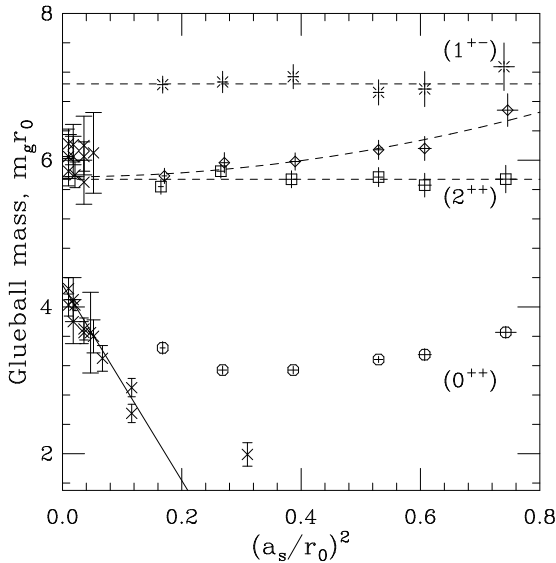}
\vspace{-8ex}
\caption{Scaling behaviour of the glueball states. The lattice irreps
$A_1^{++}, E^{++}, T_2^{++}\mbox{ and }T_1^{+-}$ are labelled $\circ, \Box,
\Diamond \mbox{ and }*$, respectively. 
Crosses indicate Wilson action data from
\protect\cite{Glueballs}. The solid line is a fit to the scalar glueball Wilson 
data, the dashed lines
are fits to the leading scaling behaviour from the improved action (see Table
\protect\ref{tab:Continuum}).\label{glueball_scale}
}
\end{figure}

Simulation data for the tensor $(2^{++})$ and pseudovector $(1^{+-})$ states
were also examined. The $\beta=2.4$
effective masses for the three lattice irreps (the $T_2^{++}$ and $E^{++}$
for the tensor and the $T_1^{+-}$ for the pseudovector) with these continuum 
quantum numbers are shown in Figure \ref{tens_vec_eff_mass}. Again, reliable 
single exponential fits
to these correlators can be performed over a range of time-slices. The scaling
behaviours of these states are included in Figure \ref{glueball_scale}. For the 
$E^{++}$ and $T_1^{+-}$ channels, no 
scaling violations are seen while the $T_2^{++}$ mass does show some cut-off
dependence. At the largest lattice spacing, this leads to a $15\%$ split in the
masses, caused by rotational invariance breaking. The cut-off dependence is 
consistent with the anticipated leading 
discretisation error, proportional to $a_s^4$ (however other functional forms
can not be ruled out). Extrapolations to the continuum limit are summarised in
Table \ref{tab:Continuum}. Here, the simplest functional form that gives a
reasonable $\chi^2$ is used in the extrapolations. Note that in the continuum
limit, the masses of the two tensor irreps are in excellent agreement, 
suggesting restoration of rotational symmetry. 

\section{CONCLUSIONS}
The simulations discussed here clearly demonstrate the advantages of using
anisotropic actions to study the glueball spectrum of QCD. Mean link
improvement was crucial in setting the couplings in the action. Using only
workstations and an improved action with anisotropy $1/3$, clear plateaux for 
the scalar,
tensor and pseudovector glueball states have been resolved, a possibility 
previously reserved only for large-scale, supercomputer-based calculations. 
Including all six simulations, a total of approximately $5\times10^9$ link 
updates
were performed, two to three orders of magnitude fewer than Wilson action 
calculations of similar statistical accuracy.

Except for the scalar glueball, all the states demonstrate the expected scaling 
behaviour for this action. 
Using $1/r_0=372$ MeV to set the
scale gives a tensor mass of $2140 \pm 45$ MeV (from the $E^{++}$ irrep) and a 
pseudovector mass
of $2620 \pm 60$ MeV. The errors quoted are purely statistical. 
In the case of the scalar, the improvement has
reduced the cutoff contamination significantly.

Alternative improved actions which
could have less contamination from the adjoint/fundamental fixed point are
under investigation. With the success in extracting the continuum masses of the 
three states considered, an extended calculation to compute the masses of the
lightest states in all 20 lattice irreps seems feasible, perhaps using a higher
anisotropy. 

We would like to thank S.J.~Dong and K.F.~Liu for use of their SU(3)
projection code. MP is grateful for financial support from UKCCS.

\begin{table}[t]
\caption{Tensor and pseudovector continuum extrapolations. The data are fit to 
one of two alternatives: (A) $m_g(a_s) r_0  = c$, or (B) $m_g(a_s) r_0 = c + d 
a_s^4$.  }
\label{tab:Continuum}
\begin{center}
\begin{tabular}{lccc}
\hline
 & Fit type & $\chi^2/$dof. & $m_gr_0(\mbox{continuum})$ \\
\hline
 $E^{++}$   & A & 0.52 & $5.74 \pm 0.12$ \\
 $T_2^{++}$ & B & 0.38 & $5.77 \pm 0.08$  \\
\hline
 $T_1^{+-}$ & A & 0.29 & $7.04 \pm 0.17$ \\
\hline
\end{tabular}
\end{center}
\end{table}


\begin{thebibliography}{9}
\bibitem{Lepage95} See for example G.P~Lepage, Nucl. Phys. {\bf B}
(Proc. Suppl.) 47 (1996) 3.
\bibitem{Morningstar95} C.J.~Morningstar and M.J.~Peardon, Nucl. Phys. {\bf B}
(Proc. Suppl.) 47 (1996) 258.
\bibitem{Morningstar96} Colin Morningstar, these proceedings; M.~Alford {\it
et.al.} to appear.
\bibitem{Lepage93} G.P.~Lepage and P.B.~Mackenzie, Phys. Rev. {\bf D48} (1993) 
2250.
\bibitem{Luescher86} M.~L\"uscher, Comm. Math. Phys. {\bf 104} (1986), 177.
\bibitem{Sommer94} R.~Sommer, Nucl. Phys. {\bf B411} (1994) 839-854.
\bibitem{FuzzSmear} APE Collaboration, Phys. Lett. {\bf B192} (1987) 163;
M.~Teper, Phys. Lett. {\bf 183B} (1986) 345.
\bibitem{Glueballs} C.~Michael and M.~Teper, Nucl. Phys. {\bf B314} (1989) 347;
J.~Sexton {\it et.al.}, Phys. Rev. Lett {\bf 75} (1995) 4563; Ph.~de Forcrand
{\it et.al.}, Phys Lett {\bf B152} (1985) 107; UKQCD Collaboration, Phys. Lett.
{\bf B309} (1993) 378. 
\bibitem{Heller96} U.~Heller,  Nucl. Phys. {\bf B}
(Proc. Suppl.) 47 (1996) 262.
\end{thebibliography}
\end{document}